\documentclass[a4paper,11pt]{article}
\usepackage{times}
\usepackage{amsmath,amssymb,amsfonts,amsthm}
\def\be{\begin{equation}}
\def\ee{\end{equation}}
\def\lb{\label}
\def\la{\langle}
\def\ra{\rangle}
\def\Sig{\Sigma}
\def\ndot{\dot{n}}
\def\D{\mbox{D}}
\newcommand{\dd}{{\rm d}}

\begin{document}
\title{\sc On the Trace-Free Einstein Equations as a Viable
Alternative to General Relativity}
\author{
{\sc George F. R. Ellis}${}^{1}$\thanks{E--mail: {\tt
George.Ellis@uct.ac.za}},
{\sc Henk van Elst}${}^{1,2}$\thanks{E--mail: {\tt
hvanelst@karlshochschule.de}},
{\sc Jeff Murugan}${}^{1,4}$\thanks{E--mail: {\tt
jeff@nassp.uct.ac.za}},
and
{\sc Jean-Philippe Uzan}${}^{1,3,4}$\thanks{E--mail: {\tt
uzan@iap.fr}}\\
{\small\em ${}^{1}$Astrophysics, Cosmology and Gravitation
Centre}\\
{\small\em Department of Mathematics and Applied Mathematics,
University of Cape Town} \\
{\small\em Rondebosch 7701, South Africa} \\
{\small\em ${}^{2}$Fakult\"{a}t I: Betriebswirtschaft und
Management, Karlshochschule International University} \\
{\small\em Karlstra\ss e 36--38, 76133 Karlsruhe, Germany}\\
{\small\em ${}^{3}$ Institut d'Astrophysique de Paris,
UMR-7095 du CNRS, Universit\'e Paris VI Pierre et Marie Curie}\\
{\small\em  98 bis bd Arago, 75014 Paris, France}\\
{\small\em ${}^{4}$ National Institute for Theoretical Physics
(NITheP), Stellenbosch 7600, South Africa}
}

\date{\normalsize{April 2, 2011}}
\maketitle
\begin{abstract}
The quantum field theoretic prediction for the vacuum energy density
leads to a value for the effective cosmological constant that is
incorrect by between 60 to 120 orders of magnitude. We review an old
proposal of replacing Einstein's Field Equations by their trace-free
part (the Trace-Free Einstein Equations), together with an
independent assumption of energy--momentum conservation by matter
fields. While this does not solve the fundamental issue of why the
cosmological constant has the value that is observed cosmologically,
it is indeed a viable theory that resolves the problem of the
discrepancy between the vacuum energy density and the observed value
of the cosmological constant. However, one has to check that, as
well as preserving the standard cosmological equations,  this does
not destroy other predictions, such as the junction conditions that
underlie the use of standard stellar models. We confirm that no
problems arise here: hence, the Trace-Free Einstein Equations are
indeed viable for cosmological and astrophysical applications.

\end{abstract}
\section{Introduction}
The interpretation of dark energy is a major puzzle \cite{Uza10}.
The gravitational effect of the quantum vacuum is expected to be
equivalent to an effective cosmological constant~\cite{zeldo}, which
according to the standard view will cause an accelerated expansion
of the universe. However, simple estimates of its expected magnitude
are very large, exceeding the observed value by between 60 and 120
orders of magnitude \cite{Wei89,Car01,FriTurHut08}; a blatant
contradiction with observations. This indicates a profound
discrepancy between General Relativity (GR) and Quantum Field Theory
(QFT) --- a major problem for theoretical
physics~\cite{pauli,Wei89}. This will presumably be resolved when a
full-blown theory of quantum gravity is finalized and accepted.
However, the field theory view of gravity as a massless spin-2
field, and with the quantum vacuum contributing to the cosmological
constant, is a half-way house between GR and a full quantum gravity
theory (which will need to have both GR and spin-2 QFT as
appropriate limits). We need to resolve that discrepancy, no matter
what final quantum gravity theory is adopted.

The key issue is: ``{\em What are the gravitational effects of the
vacuum energy?\/}'' One of the proposals for solving the problem,
summarized by Weinberg \cite{Wei89}, is use of Trace-Free Einstein
Equations (TFE) instead of the standard Einstein's Field Equations
\cite{Einstein:1915ca,Einstein:1917ce}. If these trace-free
equations are adopted, the vacuum energy has no gravitational
effect. This does not determine a unique value for the effective
cosmological constant, but it does solve the huge discrepancy
between theory and observation. However, that resolution is
dependent on this form of the field equations giving the same
results as the standard equations, not only for the solar system
and black holes (which will follow because the vacuum field
equations are unchanged), but also on the one hand for
cosmological models, and on the other for astrophysical objects
such as stellar models, where one has standard junction conditions
between the interior and exterior solutions. The former must work
out as expected (because of Weinberg's results), but how this
happens is not immediately obvious. In the latter case, it is not
{\it a priori} clear that the usual results should remain true. Indeed,
one might even expect a discrepancy between the stellar mass and the mass
of the exterior Schwarzschild solution.

In this article, we confirm that the trace-free Einstein equations are compatible with either case,
and so are indeed a viable alternative to the GR field equations
--- with the improvement that, unlike Einstein's Field Equations,
they do not suffer from the crucial discrepancy with the standard
results of quantum field theory, which imply a very large value for
the vacuum energy: so large that even solar system results would be
affected. Pauli in the 1930's (as quoted by Straumann: \cite{pauli},
pp.6-7) estimated the zero-point energy of the radiation field,
assuming that the cut-off of the theory was the electron mass. He
then estimated that the radius of the de Sitter universe ($\Lambda
\simeq 1/R^2$) would be smaller than the Earth--Moon distance;
actually, one can easily compute that $R \simeq 31~{\rm km}$
\cite{pauli}. So this value drastically affects the Solar system,
since there would be no solar system. It seems prudent to look for a
way out.

\section{Basic problem}
Classical gravitational dynamics is encoded in the Einstein Field Equations
(EFE) \cite{Einstein:1915ca,Einstein:1917ce}
\begin{equation}
\label{EFE}
G_{ab} + \Lambda g_{ab} = R_{ab} - \frac{1}{2}\,Rg_{ab}
+ \Lambda g_{ab}= \frac{8\pi G}{c^{4}}\,T_{ab} \ ,
\end{equation}
where $\Lambda$ is the cosmological constant. These are subject to
the conservation equations
\begin{equation}
\label{EFE_div}
\nabla_{b}G^{ab} = 0 \quad\Rightarrow\quad
\nabla_{b}T^{ab}=0 \ ,
\end{equation}
for the total energy--momentum tensor,
which guarantee the consistency of the time development of the EFE.

In the standard cosmological application, the metric tensor is
assumed to take the spatially homogeneous and isotropic form
\begin{equation}
\label{RW metric} ds^2 = -\,c^{2}dt^2 + a^2(t)d\sigma^2 \ ,
\qquad
u^a = \delta_{0}{}^{a} \ ,
\end{equation}
with $a(t)$ a universal time-dependent scale factor, $d\sigma^2$ the
metric of a 3-space of constant curvature $k$, and $u^{a}$ the
normalised matter 4-velocity field ($u_{a}u^{a} = -1$). Because of
the symmetries of the metric, the energy--momentum tensor of
the matter necessarily takes a perfect fluid form:
\begin{equation}
\label{PF}
T_{ab} = \left(\rho + \frac{p}{c^2}\right)c^{2}u_a u_b
+ p\,g_{ab} \ ,
\end{equation}
where the matter mass density $\rho$ and the matter isotropic
pressure $p$  are related through an equation of state $p =
p(\rho)$. Whenever the energy--momentum tensor takes the form
(\ref{PF}), the conservation equations (\ref{EFE_div}) give the
mass--energy conservation relation
\begin{equation}
\label{cons1}
\dot{\rho} + 3\,\frac{\dot{a}}{a}\left(\rho+\frac{p}{c^2}\right)=0 \ ,
\end{equation}
where a dot denotes a covariant derivative with respect to proper
time, $d/dt$. For the case of the Robertson--Walker metric (\ref{RW
metric}), the EFE (\ref{EFE}) reduce to two non-trivial equations:
the Raychaudhuri equation
\begin{equation}
\label{ray}
3\,\frac{\ddot{a}}{a} + 4\pi G\left(\rho+3\,\frac{p}{c^2}\right)
- \Lambda c^{2} = 0 \ ,
\end{equation}
and the Friedmann equation
\begin{equation}
\label{fried}
3\left(\frac{\dot{a}}{a}\right)^{2} = -\,\frac{3kc^{2}}{a^2} + 8\pi G\rho
+\Lambda c^{2} \ .
\end{equation}
The latter equation is a first
integral of the other two; indeed any two of
(\ref{cons1}) to (\ref{fried}) imply the third. When they are satisfied,
all 10 EFE (\ref{EFE}) are satisfied.

The key equation as regards gravitational attraction is the
Raychaudhuri equation (\ref{ray}), which shows that $(\rho +
3p/c^2)$ is the active gravitational mass density. The equation of
state for a vacuum
\begin{equation}
\label{vac}
p_{\rm vac} = -\,\rho_{\rm vac}c^2
\end{equation}
shows this is negative:
\begin{equation}
\label{vac1}
\left(\rho_{\rm vac}+3\,\frac{p_{\rm vac}}{c^2}\right)
= -\,2\rho_{\rm vac} \ ,
\end{equation}
where we can represent the effect of the vacuum in (\ref{ray})
either as a fluid with equation of state (\ref{vac}), or as an
effective cosmological constant,
\begin{equation}
\Lambda_{\rm vac} =  \frac{8\pi G}{c^{2}}\,\rho_{\rm vac} \ .
\end{equation}

The problem is that the QFT view of the vacuum as an infinite set of
oscillators, each with zero-point oscillatory energy $\frac{1}{2}
\hbar \omega_k$, gives a diverging value for the vacuum energy
$E_{\rm vac}$. With a suitable high-energy cut-off, the vacuum
energy density is estimated by Weinberg \cite{Wei89} to be of the
order
\begin{equation}
\label{vac2}
\la\rho_{\rm vac}\ra\,\simeq \,2 \times 10^{71}\,{\rm GeV}^4 \ ,
\end{equation}
whereas the effective value of the cosmological constant as
determined by astronomical observations is of the order
\begin{equation}
\label{vac3}
\la\rho_{\rm obs}\ra\,\simeq\, 10^{-47}\,{\rm GeV}^4 \ .
\end{equation}
Because the value (\ref{vac2}) is a constant, it has no effect
on local dynamics, and so can be subtracted off the total energy
density as far as local physics is concerned, but on the standard
view, because of (\ref{vac1}), it will have a gravitational effect
hugely bigger than the observed value (\ref{vac3}).

There are three ways out of this problem.
\begin{enumerate}
  \item Other fields may contribute negative energy densities
  that will cancel the positive terms, and leave a very small
  residue as observed; or maybe there is a symmetry implying
  $\rho_{\rm vac} + \Lambda_{\rm bare} = 0$,
  and one has to have extra fields (`quintessence') to give
  the observed acceleration. This is indeed possible in
  principle; for example, each mode of a Dirac field gives a
  negative contribution to the vacuum energy (hence, this is
  the option that would be realised via supersymmetry, were
  supersymmetry not broken in the real universe). But it is
  very hard to make this work in practice: many fields
  contribute to the vacuum energy density, and it is highly
  unlikely they would just happen to cancel the value (\ref{vac2})
  accurately to 120 decimal places, but not exactly, so as
  to give (\ref{vac3}).
  \item The value in (\ref{vac2}) is only an expectation value:
    maybe we live in a multiverse where $\rho_{\rm vac}$ takes all
    sorts of values, and anthropic selection effects determine we
    live in a universe domain where its value (\ref{vac3}) is very
    far from the expectation value. This is a possible solution,
    but many regard it as an act of desperation, abandoning the
    idea of $\Lambda$ as a fundamental constant in favour of
    regarding its value as being contingent.
  \item Maybe the EFE (\ref{EFE}) are not the true effective
  equations of gravitational interactions: a variant to the EFE
  arises from the underlying quantum gravity theory, and negates
  the gravitational effect of the vacuum.
\end{enumerate}
We explore aspects of the latter option in this article.

\section{Trace-free Einstein gravity}
Einstein continually worried about what should go on the right-hand
side of the relativistic gravitational field equations. An
interesting proposal is to take the trace-free part of the EFE to
get the Trace-Free Einstein Equations (TFE), which is a subset of
the EFE that can give back the full EFE with an integration
constant. This is an old proposal, essentially initiated by Einstein
\cite{Ein19} himself, and developed by many others since, see
\cite[pp.~11--13]{Wei89} and references given there. More recently,
it has been developed under the name of ``unimodular gravity'', see
\cite{AndFin71,Unr89,FinGalBau00,Smo09} and references therein.

We use a hat to denote the trace-free part of a symmetric tensor:
so
$$\hat{G}_{ab} := R_{ab} - \frac{1}{4}\,R g_{ab} \ ,
\quad
\hat{T}_{ab} := T_{ab} - \frac{1}{4}\,T g_{ab}
\quad\Rightarrow\quad
\hat{G}_{a}{}^a = 0 \ ,
\quad
\hat{T}_{a}{}^a = 0 \ .$$
On taking its trace-free part, the EFE
(\ref{EFE}) implies the TFE
\begin{equation}
\label{EFE_TF1}
\hat{G}_{ab} = \frac{8\pi G}{c^{4}}\,\hat{T}_{ab}
\quad\Leftrightarrow\quad R_{ab} - \frac{1}{4}\,R g_{ab} =
\frac{8\pi G}{c^{4}} \left(T_{ab} - \frac{1}{4}\,T g_{ab}\right) \ .
\end{equation}
We adopt these as the gravitational field equations, instead of
(\ref{EFE}). The twice-contracted second Bianchi identity
for the Einstein tensor $G_{ab}$ still holds:
\begin{equation}
\label{EFE_div1}
\nabla_{b}G^{ab} = 0 \ .
\end{equation}
However, now the corresponding divergence relation for the
energy--momentum tensor $T_{ab}$,
\begin{equation}
\label{EFE_div2}
\nabla_{b}T^{ab} = 0 \ ,
\end{equation}
is no longer a consequence of the geometrical identity
(\ref{EFE_div1}), as in (\ref{EFE_div}): it is a
{\em separate assumption\/}.

The gravity theory based on the TFE recovers all the vacuum
solutions of the EFE unchanged, so e.g. results from the
Schwarzschild and Kerr black hole solutions are still valid.
However, it no longer has a
cosmological constant problem, as $\Lambda$ does not affect
spacetime curvature. Indeed, for a perfect fluid the matter
source term is the manifestly trace-free energy--momentum
tensor
\begin{equation}
\label{EFE mat1}
\hat{T}_{ab} = \left(\rho+\frac{p}{c^2}\right)c^{2}
\left(u_a u_b+\frac{1}{4}\,g_{ab}\right) \ ;
\end{equation}
hence, matter enters the field equations only in terms of the
inertial mass density $(\rho+p/c^2)$, which vanishes for a vacuum.
However, as discussed in \cite{Wei89}, the
theory acquires a new {\em integrability condition\/}:
differentiating (\ref{EFE_TF1}), and using (\ref{EFE_div1}) and
(\ref{EFE_div2}), gives
\begin{equation}
\label{EFE_TF2}
\nabla_{b}\left(R^{ab} - \frac{1}{4}\,R g^{ab}\right)
= \frac{8\pi G}{c^{4}}\,\nabla_{b}\left(T^{ab}
- \frac{1}{4}\,T g^{ab}\right)
\quad\Rightarrow\quad
\nabla_{a}R = -\,\frac{8\pi G}{c^{4}}\,\nabla_{a}T \ .
\end{equation}
Integrating, $\displaystyle\left(R+\frac{8\pi G}{c^{4}}\,
T\right)$ is a constant:
\begin{equation}
\label{EFE_TF3}
R+\frac{8\pi G}{c^{4}}\,T =: 4\hat{\Lambda} \ .
\end{equation}
Substituting into (\ref{EFE_TF1}) to eliminate $T$ gives
back (\ref{EFE}), but with a
new effective cosmological constant:
\begin{equation}
\label{EFE3}
R_{ab} - \frac{1}{2}\,R g_{ab} + \hat{\Lambda}g_{ab}
= \frac{8\pi G}{c^{4}}\,T_{ab} \ .
\end{equation}
So the way it works is as follows: we {\em assume\/} both the TFE
(\ref{EFE_TF1}) and the matter conservation equations
(\ref{EFE_div2}). The integrability condition (\ref{EFE_TF2})
follows from these equations. Integrating gives (\ref{EFE3}). This
is not surprising in that we have assumed the validity of the
trace-free part of the EFE (\ref{EFE}); and (\ref{EFE_TF3}) is
just the trace of those equations. The same consistency results
hold as for the full EFE: by (\ref{EFE_div1}) the time evolution
equations amongst the TFE preserve the constraints. This result
can be implemented as a dynamical theory where the spacetime volume
density $\sqrt{-g}$ is {\em not\/} a dynamical
variable \cite{Wei89}.

Hence, we have a remarkable result \cite{Wei89}: \emph{the TFE
(\ref{EFE_TF1}) together with the differential relations
(\ref{EFE_div2}) are functionally equivalent to the EFE
(\ref{EFE3}), with the cosmological constant an arbitrary
integration constant $\hat{\Lambda}$ unrelated to the vacuum energy
$\Lambda_{\rm vac}$.} Note this is not the same as either
\begin{equation}
\label{EFE_TF4}
\hat{G}_{ab} = \frac{8\pi G}{c^{4}}\,T_{ab}
\quad\Rightarrow\quad T = 0 \ , \quad R = \text{constant}
\end{equation}
(as proposed by Einstein in 1919, see \cite{Ein19}), or
\begin{equation}
\label{EFE_TF5}
G_{ab} = \frac{8\pi G}{c^{4}}\,\hat{T}_{ab}
\quad\Rightarrow\quad R = 0 \ , \quad T = \text{constant}
\end{equation}
(which also decouples spacetime curvature from vacuum energy). Each
of these proposals has 10 field equations, and allows only very
restricted forms of matter to occur. The proposal that makes
physical sense is (\ref{EFE_TF1}), where both sides of the equation
have been specifically constructed so as to have the same symmetry.
Then we have 9 field equations and the energy--momentum tensor
of the matter can have a non-zero trace, but only the trace-free
part gravitates; this can accommodate generic inhomogeneous matter.

What about experiments? \emph{The experimental predictions of the
two theories are the same, so no experiment can tell the difference
between them, except for one fundamental feature: the EFE (confirmed
in the solar system and by binary pulsar measurements to high
accuracy) together with the QFT prediction for the vacuum energy
density (confirmed by Casimir force measurements) give the wrong
answer by many orders of magnitude; the TFE does not suffer this
problem}. In this respect, the TFE are strongly preferred by
experiment.

This trace-free Einstein gravity theory then should give cosmology
without the vacuum energy problem. We explore this in the next
section, but we will not deal with the relation of this theory to
commonplace variational principle approaches, which is adequately
covered elsewhere (see \cite{AndFin71,Wei89,Smo09} and references
there).

\section{FLRW cosmologies}
This works out in the spatially homogeneous and isotropic case as a
special case of the above general theory (as, of course, it has to).
That is, starting from (\ref{cons1}), (\ref{ray}) and (\ref{fried}),
we first determine a trace-free dynamic equation by eliminating
$\Lambda$ between (\ref{ray}) and (\ref{fried}), and then derive an
integrability condition for this equation. Thereafter, we show how
this latter equation can be integrated to recover (\ref{ray}) and
(\ref{fried}) with a new effective cosmological constant
$\hat{\Lambda}$ as an integration constant, independent of any
dynamical values we might assign to the vacuum energy. Thus, this
solves the major problem of a vacuum energy many orders of magnitude
larger than measured by cosmological observations.

In detail: the trace-free equation is
\begin{equation}
\label{FL_TF}
\frac{\ddot{a}}{a} - \left(\frac{\dot{a}}{a}\right)^{2}
- \frac{kc^{2}}{a^{2}}
= -\,4\pi G\left(\rho+\frac{p}{c^{2}}\right) \ .
\end{equation}
The spacetime Ricci curvature scalar $R$ is given by
\begin{equation}
\label{Ricci scalar}
Rc^{2} = -\,\frac{8\pi G}{c^{2}}\,T + 4\Lambda c^{2}
= 8\pi G\left(\rho-3\,\frac{p}{c^2}\right) + 4\Lambda c^{2}
= 6\,\left[\,\frac{\ddot{a}}{a}+\left(\frac{\dot{a}}{a}\right)^{2}
+\frac{kc^{2}}{a^2}\,\right] \ .
\end{equation}
The time derivative of this equation is the needed integrability
condition for (\ref{FL_TF}), encoding the fact that $\Lambda$ is a
constant (see (\ref{EFE_TF2})). It has the form
\begin{equation}
\label{TF_integ}
0 = 6\,\frac{d}{dt}\left[\,\frac{\ddot{a}}{a}
+\left(\frac{\dot{a}}{a}\right)^{2}
+\frac{kc^{2}}{a^2}
-\frac{4\pi G}{3}\left(\rho-3\,\frac{p}{c^{2}}\right)\,\right] \ .
\end{equation}
Now (\ref{TF_integ}), being a vanishing total time
derivative, can be easily integrated to yield an integration
constant $\displaystyle \frac{2}{3}\,\hat{\Lambda}$. Eliminating
$\ddot{a}/a$ between (\ref{FL_TF}) and the integral of
(\ref{TF_integ}), and including the (now assumed) mass--energy
conservation relation (\ref{cons1}), one recovers the original dynamic
equations (\ref{ray}) and (\ref{fried}), but with a renormalized
$\Lambda$. That is, we obtain in this way the Raychaudhuri equation
\begin{equation}
\label{ray4}
3\,\frac{\ddot{a}}{a} + 4\pi G\left(\rho + 3\,\frac{p}{c^2}\right)
+ \hat{\Lambda}c^{2} = 0 \ ,
\end{equation}
and the Friedmann equation
\begin{equation}
\label{fried3}
3\left(\frac{\dot{a}}{a}\right)^{2}
= -\,\frac{3kc^{2}}{a^2}+ 8\pi G\rho
+ \hat{\Lambda}c^{2} \ ,
\end{equation}
with effective cosmological constant $\hat{\Lambda}$ --- an
arbitrary integration constant, unrelated to $\Lambda_{\rm vac}$.
This solves the basic discrepancy between QFT
estimates of the energy density of the quantum vacuum, and the
disastrous result if we assume this is a source term in the
Raychaudhuri equation in an obvious way. Thus, we arrive at

\medskip
\noindent \textbf{Hypothesis}: The EFE are not the true effective
equations of gravity: rather --- whatever the underlying quantum
theory of gravity --- the effective theory is trace-free Einstein
gravity, as described above.

\medskip
\noindent In that case the basic equations of cosmology are
(\ref{FL_TF}) and (\ref{cons1}), supplemented by an equation of
state determining the pressure from the mass density, and
({\ref{ray4}) and ({\ref{fried3}) are consequences. The value of the
vacuum term $\Lambda_{\rm vac}$ does not affect the cosmic
expansion. Note the following remarkable feature: \emph{the active
gravitational mass density $\rho_{\rm grav} = (\rho +3p/c^2)$ does
{\em not\/} occur in either of the equations (\ref{FL_TF}) or
(\ref{cons1}) that we take as the basis of the dynamics: only the
inertial mass density $\rho_{\rm inert}= (\rho+p/c^2)$ occurs there.
Nevertheless, in the end $\rho_{\rm grav}$ turns out to be the
effective gravitational mass density}, see (\ref{ray4}). What
happens is that while (\ref{cons1}) determines the time evolution of
the energy density, the assumed equation of state determines the
time evolution of the pressure, and does so in such a way that
(\ref{ray4}) results. It is interesting to note that (\ref{FL_TF})
is in the form of the dynamical equations often used for studies of
inflationary universe models.\footnote{We thank Roy Maartens for
this comment.}

Let us also mention an important point concerning scalar
field cosmology. The transformation
$T_{ab} \rightarrow T_{ab}+\Lambda g_{ab}$ leaves
$\hat T_{ab}$ and the conservation equation~(\ref{EFE_div2})
unchanged only if $\Lambda$ is a constant. In the case of a
self-interacting scalar field $\varphi$ evolving in a
potential $V(\varphi)$, it is easily checked for the
energy--momentum tensor that
\begin{equation}
T_{ab}=\partial_{a}\varphi\partial_{b}\varphi
-\frac{1}{2}\left[\,\partial_{c}\varphi\partial^{c}\varphi
+2V(\varphi)\,\right]g_{ab}
\quad\Rightarrow\quad
\hat{T}_{ab}=\partial_{a}\varphi\partial_{b}\varphi
-\frac{1}{4}(\partial_{c}\varphi\partial^{c}\varphi)g_{ab}
\end{equation}
whatever the potential, the latter of which, therefore, does
not appear in the TFE~(\ref{EFE_TF2}), and so has no
gravitational effect. Nevertheless, the potential is, of course,
of dynamical relevance, since it cannot be eliminated from the
conservation equation as long as $\dot\varphi\not=0$. For an
FLRW scalar field cosmology, (\ref{EFE_div2}) yields the
equation of motion
\begin{equation}
\dot\varphi\left(\ddot\varphi+3H\dot\varphi
+ \frac{dV}{d\varphi} \right)= 0 \ .
\end{equation}
The dynamics of a universe containing a scalar field (e.g. during
inflation) will then be the same as in standard GR. Note, however,
that we now have the freedom to shift the minimum of the potential
at will, since $V(\varphi) \rightarrow V(\varphi)+V_{0}$ leaves
the equation of motion unchanged.

\section{Junction conditions}
To complete the picture of the trace-free Einstein gravity
proposal, we need to verify that in the case of a stellar model
the interior and exterior solutions match, so that the mass
measured outside is the same as the mass of the interior solution;
if this was not true, it would be a disaster for these equations,
as they would not give standard results for stellar structure
models. In the GR context, this issue is settled by the
Darmois--Israel junction conditions \cite{dar1927,isr1966},
which arise from the local embedding relations of Gau\ss\ and
Codazzi for a timelike or spacelike 3-surface in a 4-dimensional
spacetime manifold.

\subsection{Covariant $3+1$ viewpoint}
Consider a non-null 3-surface $\Sig:\{\phi=\text{constant}\}$
(of codimension 1) embedded in a spacetime manifold, with unit
normal $n_{a}$ given by
\be
n_{a}:=\nabla_{a}\phi \ ,
\ee
such that
\be
n_{a}n^{a}=\varepsilon=\begin{cases}
+1 \ ,& \text{when}\ \Sig\ \text{is\ timelike}\\
-1 \ ,& \text{when}\ \Sig\ \text{is\ spacelike}
\end{cases}
\ee
applies. Then
\be
\delta_{a}{}^{b} =: \varepsilon n_{a}n^{b} + h_{a}{}^{b}
\ee
defines tensors $\perp_{a}{}^{b}
:= \varepsilon n_{a}n^{b}$ and $h_{a}{}^{b}$
which project orthogonally and tangentially to $\Sig$,
respectively, with $0=h_{a}{}^{b}n_{b}$. The covariant
derivative of $n_{a}$ is given by
\be
\lb{ncovder}
\nabla_{a}n_{b} = \varepsilon n_{a}\ndot_{b} - K_{ab} \ ,
\ee
which defines the non-geodesity of $n_{a}$
by $\ndot_{a}:=n^{b}\nabla_{b}n^{a}$,
and the extrinsic curvature tensor of $\Sig$ by
$K_{ab}:=-h_{a}{}^{c}h_{b}{}^{d}\nabla_{c}n_{d}=:-\D_{(a}n_{b)}$,
where $\D_{a}$ is the induced connection of $\Sig$. Note that
$0=\ndot_{a}n^{a}=K_{ab}n^{b}$ applies.

Certain orthogonal and tangential projections
(by means of $\perp_{a}{}^{b}$ and $h_{a}{}^{b}$) of the
4-dimensional Ricci identity generate for $\Sig$ the
well-known local embedding equations
of Gau\ss\ and Codazzi (see e.g. \cite{ste1991}),
\begin{eqnarray}
\lb{4dricidtttt}
{}^{3}\!R_{abcd}
- \varepsilon(K_{ac}K_{bd}-K_{ad}K_{bc})
& = & h_{a}{}^{e}h_{b}{}^{f}h_{c}{}^{g}h_{d}{}^{h}R_{efgh} \\
\lb{4dricidtttn}
\D_{a}K_{bc} - \D_{b}K_{ac}
& = & -h_{a}{}^{d}h_{b}{}^{e}h_{c}{}^{f}n^{g}R_{defg} \ ,
\end{eqnarray}
with ${}^{3}\!R_{abcd}$ the intrinsic 3-Riemann curvature
tensor of $\Sig$. Linear combinations of contractions of
these relations then lead to (see \cite{mtw,ste1991})
\begin{eqnarray}
\lb{einnn}
n^{a}n^{b}G_{ab} & = & \frac{1}{2}\left(K^{2}-K_{ab}K^{ab}
-\varepsilon\,{}^{3}\!R\right) \\
\lb{einnt}
-n^{b}h_{a}{}^{c}G_{bc} & = & \D_{b}K_{a}{}^{b} - \D_{a}K \ ,
\end{eqnarray}
where $K:=K_{a}{}^{a}$. In the absence of matter surface layers,
Darmois' junction conditions \cite{dar1927} require
that across $\Sigma$ the induced 3-metric and the extrinsic curvature
be continuous, i.e.,
\begin{eqnarray}
0 & = & [h_{ab}] \quad\Rightarrow\quad
0 = [\D_{a}] \ , \quad
0 = [{}^{3}\!R_{abcd}] \ , \\
0 & = & [K_{ab}] \quad\Rightarrow\quad
0 = [\D_{a}K_{bc}] \ ,\label{eq38}
\end{eqnarray}
employing $[Q]:=Q_{+}-Q_{-}$ to denote the change of any
quantity $Q$ across $\Sigma$. With (\ref{einnn}) and
(\ref{einnt}), this leads to Israel's junction
conditions \cite{isr1966}
\be
\lb{efejunccon}
0 = [n_{a}n_{b}G^{ab}] \ , \qquad
0 = [n_{b}h_{c}{}^{a}G^{bc}] \ .
\ee
As this is a purely geometrical result, it does not make a
qualitative difference whether we use (i) the EFE (\ref{EFE}),
or (ii) the TFE (\ref{EFE_TF1}) together with (\ref{EFE_TF3}),
to replace the Einstein tensor $G_{ab}$ in these relations.

\subsection{Including a matter surface layer}
The previous consideration of the Darmois--Israel junction
conditions assumes that there is no matter surface layer on
the 3-surface $\Sigma$ between the two parts of spacetime to
be matched. An alternative approach to the derivation of the
junction conditions relies on the use of a local coordinate
approach. Using the same conventions as in the previous paragraph,
in a parametric local representation of $\Sig$ given by
$x^{a}=X^{a}(\sigma^{i})$ (where $a,b, \ldots = 0,\ldots, 3$ and
$i,j, \ldots = 1,\ldots, 3$) the coordinate components of
the induced 3-metric are expressed by
\begin{equation}
\gamma_{ij}:=\frac{\partial X^{a}}{\partial\sigma^{i}}
\frac{\partial X^{b}}{\partial\sigma^{j}}g_{ab} \ .
\end{equation}
Viewed from the 4-dimensional perspective, this is the
first fundamental form
$\displaystyle h^{ab}=\frac{\partial X^{a}}{\partial\sigma^{i}}
\frac{\partial X^{b}}{\partial\sigma^{j}}\gamma^{ij}
= g^{ab}-\varepsilon n^{a}n^{b}$.
If we assume that $n_a$ is smoothly continued into the 4-dimensional
spacetime along a set geodesics, i.e. such that $\dot{n}_{a}=0$,
the extrinsic curvature reduces to $K_{ab} = -\nabla_a n_b$
[see (\ref{ncovder})]. While $g_{ab}$ is everywhere continuous,
its derivative along $n_{a}$ may not be continuous across $\Sigma$,
so that $K_{ab}$ may also be discontinuous.

Now consider a neighbourhood of $\Sigma$ represented in terms of
a Gau\ss ian normal coordinate system, where the coordinate $s$
varying in the interval $s_{-}<s<s_{+}$ measures the
distance to $\Sigma$; we assume $\Sigma$ to be located at $s=0$.
This implies that $n_a=\nabla_a s$. We then denote by
$Q^{\prime} =\partial_{s}Q=n^{a}\partial_{a}Q$
the partial derivative of any quantity $Q$ along the
normal direction to $\Sigma$. The change of $Q$ across
$\Sigma$ is given by
\begin{equation}
[Q]=Q_{+}-Q_{-}
=\overline Q^{\prime}
:=\int_{s_{-}}^{s_{+}}Q^{\prime} \dd s \ .
\end{equation}
This definition is such that $Q^{\prime} \rightarrow
\overline{Q}{}^{\prime}\delta(s)$ in the limit of an
infinitesimally thin 3-surface. We use angular brackets to
denote the mean value,
\begin{equation}
\langle Q\rangle:=\frac{1}{2}\left(Q_{+} + Q_{-}\right).
\end{equation}
In Gau\ss ian normal coordinates, equations~(\ref{4dricidtttt})
and~(\ref{4dricidtttn}) are complemented by the spatial
projection~\cite{mtw}
\begin{equation}
\label{4dricidttnn}
h_{a}{}^{c}h_{d}{}^{b}G_{c}{}^{d} = {}^{3}G_{a}{}^{b}
+ \varepsilon n^c\partial_c\left(K_{a}{}^{b}
- Kh_{a}{}^{b}\right) - KK_{a}{}^{b}
+ \frac{1}{2}\left(K^2+K_{cd}K^{cd}\right)h_{a}{}^{b} \ .
\end{equation}
If there is matter localised on $\Sigma$, then its
energy--momentum tensor can be computed as
\begin{equation}
S_{a}{}^{b} = \overline{T}_{a}{}^{b}
= \lim_{s\rightarrow 0} \int_{-s}^{s} T_{a}{}^{b} \dd s \ ,
\end{equation}
which satisfies $0=S_{ab}n^{b}$. In the infinitesimally
thin limit, the 4-dimensional energy--momentum tensor
on $\Sigma$ is of the form $T_{ab}=S_{ab}\delta(s)$.
For the geometry of $\Sigma$ to be well-defined, both
$g_{ab}$ and $h_{ab}$ should be continous across $\Sigma$
and thus contain neither a $\delta(s)$ factor nor a
discontinuity. The first implies that ${}^3R_{abcd}$ and
thus ${}^3R_{ab}$ have no $\delta(s)$ factor, while the
second implies that $K_{ab}$ has no $\delta(s)$ factor.
Integration of (\ref{4dricidttnn}) then implies that
\begin{equation}
h_{a}{}^{c}h_{d}{}^{b}\overline{G}_{c}{}^{d}
= \varepsilon\left[K_{a}{}^{b} - Kh_{a}{}^{b}\right] \ ,
\end{equation}
since only the terms containing a derivative along $s$
contribute to the integral, i.e.,
$G_{ab} \simeq \varepsilon(K_{ab}^{\prime}-K^{\prime}h_{ab})$.
Using that $G_{ab}=(G_{cd}n^cn^d)n_an_b
+\varepsilon G_{cd}n^ch_{a}{}^{d}n_b
+G_{cd}h_{a}{}^{c}h_{b}{}^{d}$, and the fact that the
intregrals along $n^a$ of expressions~(\ref{4dricidtttt})
and~(\ref{4dricidtttn}) vanish even when matter is localized
on $\Sigma$, we conclude that
\begin{equation}
\overline{G}_{a}{}^{b}
= \varepsilon\left[K_{a}{}^{b} - Kh_{a}{}^{b}\right] \ .
\end{equation}
With the standard EFE~(\ref{EFE}), this would imply that
$\displaystyle \left[K_{a}{}^{b} - Kh_{a}{}^{b}\right]
=\frac{8\pi G}{c^{4}}\,S_{a}{}^{b}$, a generalisation
of (\ref{eq38}).

For the TFE (\ref{EFE_TF1}), we first need to compute the
jump in the trace-free Einstein tensor $\hat{G}_{ab}$
by using that $G \simeq -2\varepsilon K^{\prime}$, so that
\begin{equation}
\label{jc1}
\overline{\hat G}_{ab} = \varepsilon\left[K_{ab}
- K\left(h_{ab}-\frac{1}{2}\,g_{ab}\right) \right]
= \frac{8\pi G}{c^{4}}\left(S_{ab}-\frac{1}{4}\,Sg_{ab}\right)
\ .
\end{equation}
While the junction condition for the trace $K$ cannot be
extracted from this equation, one has to rely on
(\ref{EFE_TF2}). We decompose the total energy--momentum
tensor according to $T_{ab}= T^{\rm mat}_{ab}
+ T^{\Sigma}_{ab}$, with $T^{\Sigma}_{ab}=S_{ab}\delta(s)$.
The hypothesis~(\ref{EFE_div2}) implies that
\begin{equation}
\nabla_{b}T^{ab}_\Sigma = f^{a} = -\nabla_{b}T_{\rm mat}^{ab} \ .
\end{equation}
We assume for simplicity that $\Sigma$ does not interact with
the matter fields (see e.g. \cite{bb} for the treatment of
such a case, and \cite{cu} for an example in which $\Sigma$
is coupled to a form field allowing a jump in the
cosmological constant). Since $\nabla_{b}T_{\rm mat}^{ab}=0$,
this implies that $n_{b}(T_{\rm mat}^{ab})^{\prime}=0$,
and thus $[T_{\rm mat}^{ab}]\perp_{ab}=0$. Then, (\ref{EFE_TF2})
implies $\displaystyle n^a\nabla_a\left(R
+\frac{8\pi G}{c^{4}}\,T\right)=0$, so that $\displaystyle R
+\frac{8\pi G}{c^{4}}\,T$ remains constant across $\Sigma$
(i.e., $\overline{\hat\Lambda}=0$). Now, since
$R \simeq \varepsilon 2K^{\prime}$, we deduce that
$\displaystyle \bar R = \varepsilon 2[K]
= -\frac{8\pi G}{c^{4}}\,S$, where the second equality arises
from the constancy of $\displaystyle R+\frac{8\pi G}{c^{4}}\,T$,
which implies $\displaystyle \bar{R}=-\frac{8\pi G}{c^{4}}\,S$.
It follows that $\displaystyle \frac{1}{2}\,\varepsilon [K]
g_{ab} = -\frac{8\pi G}{c^{4}}\,\frac{1}{4}\,Sg_{ab}$, so
that the junction condition~(\ref{jc1}) reduces to
\begin{equation}
\label{jc2}
\varepsilon\left[K_{ab} - K h_{ab} \right]
= \frac{8\pi G}{c^{4}}\,S_{ab} \ ,
\end{equation}
i.e., to the same relation as in classical GR.

In conclusion, the standard GR stellar structure models, with
an interior solution matched to an exterior solution across a
suitable 3-surface $\Sigma$, will remain valid in the case of
trace-free Einstein gravity. There will not be a problem of the
interior and exterior masses not matching. Similar issues arise
for the junction conditions in Swiss-Cheese models. Again, the
TFE will be acceptable: the usual mass matching condition will be
fulfilled, so they do not lead to anomalies here either.

\section{Viability of the TFE equations}
We have revisited the possibility that the true effective
gravitational field equations are given by the TFE, implying that
only the trace-free part of the energy--momentum tensor $T_{ab}$ of
matter is gravitating. Then the effective cosmological constant
$\hat{\Lambda}$ is a constant of integration that is arbitrarily
disposable (as in classical GR), and, hence, is independent of any
fundamental value assigned to $\Lambda$ (cf. \cite[p. 196]{Ein19}).
We do not require a fine tuning $\rho_{\rm vac} + \Lambda \simeq
\Lambda_{\rm obs}$, because $\rho_{\rm vac}$ is not gravitating.

Thus, employing the TFE in place of the EFE appears to be a good
theoretical assumption to make: any huge $\Lambda_{\rm vac}$ is
powerless to affect cosmology, or indeed the solar system, as the
zero point energy will not affect spacetime geometry. The EFE will
be as usual, but with $\hat{\Lambda}$ an integration constant that
may be small, or may be zero. As observations indicate, this
constant corresponds to a particular cosmological length scale
($\hat{\Lambda} \simeq H_0^2$) that should be determined from
initial conditions for our universe; see (\ref{EFE_TF2}). In that
sense, the vacuum energy problem vanishes in the trace-free Einstein
gravity proposal, while the almost equality between $\hat\Lambda$
and the Hubble constant, i.e., the coincidence problem, remains.

We have checked here that these equations are compatible with usual
cosmological models and also with  standard junction conditions,
without and with a matter surface layer. In particular, this
ensures that stellar structure models will be the same as in GR:
there will not be any mass anomaly between the interior and exterior
solutions. Hence, the TFE work both for cosmology and for
astrophysics.

Overall, this proposal does not solve the issue of why the
cosmological constant has the value it has today, but
it does resolve the issue of why it does not have the huge value
implied by the obvious use of the QFT prediction for the vacuum
energy in conjunction with the EFE. The patently incorrect result
obtained in this way is a major crisis for theoretical physics,
because it suggests a profound contradiction between two of our most
successful theories, namely QFT and GR. Use of the TFE instead of
the EFE solves that problem.

\section*{Acknowledgments}
We thank Roy Maartens for helpful comments on an earlier draft of
this article.


\small{Version April 2, 2011: gfre}
\end{document}